\documentclass[12pt]{article}
\usepackage{graphicx,epsfig}
\usepackage{cite}
\title{Influence of spin ordering on superconducting correlations in the
spin-one-half Falicov-Kimball model with Hund and Hubbard coupling}
\author{Pavol Farka\v sovsk\'y\\
Institute  of  Experimental  Physics,  Slovak   Academy   of
Sciences\\
Watsonova 47, 043 53 Ko\v {s}ice, Slovakia}
\date{}
\begin{document}
\baselineskip=24pt
\maketitle
\begin{abstract}
The generalized spin-one-half Falicov-Kimball model with Hund and 
Hubbard coupling is used to examine effects of spin ordering on 
superconducting correlations in the strongly correlated electron and spin 
systems. It is found that the ferromagnetic spin clusters (lines, bands, domains) 
suppress the superconducting correlations in the d-wave chanel, while 
the antiferromagnetic ones have the fully opposite effect. 
The enhancement of the superconducting correlations due to 
the antiferromagnetic spin ordering is by factor 3 in the axial striped
phase and even by the factor 8 in the phase segregated phase. 
\end{abstract}

\newpage
\section{Introduction}
The problem of formation of the charge and spin stripe order and its relation 
to superconductivity belongs surely to  one of the most exciting ideas 
of contemporary solid state physics. The reason is clearly due to the 
observation of  such an ordering in doped nickelate~\cite{Ni}, 
cuprate~\cite{Cu} and cobaltate~\cite{Co} materials, some of which 
constitute materials that exhibit high-temperature superconductivity.
Unfortunately, despite an enormous 
research activity in the past the relation between the charge/spin 
ordering and the superconductivity is still controversial 
(an excellent review of relevant works dealing with this subject 
can be found in~\cite{Oles2}). A considerable progress in this field 
has been achieved recently by Maier et al.~\cite{Maier} and  Mondaini 
et al.~\cite{Mon}. Both groups studied the two-dimensional Hubbard model,
in which stripes are introduced externally by applying a spatially 
varying local potential $V_i$, and they found a significant enhancement 
of the d-wave pairing correlations. However, it should be noted
that the potential $V_i$ is phenomenological and as such has no direct 
microscopic origin that corresponds to a degree of freedom
in the actual materials. Contrary to this approach, we have presented
very recently~\cite{epl} an alternative model of coexistence of the charge/spin 
stripe order and superconductivity in the strongly correlated systems.   
Our approach is based on a generalized spin-one-half Falicov-Kimball 
model that besides the spin-independent $U_{fd}$ as well as spin-dependent 
$J_z$ Coulomb interaction between the localized $f$ and itinerant 
$d$ electrons takes into account the Hubbard interaction  between 
$d$ ($f$) electrons of opposite spins. It is found that in the presence 
of all above mentioned interactions the model stabilizes three basic types 
of charge/spin ordering, and namely, the axial striped phases, the regular 
$n$-molecular phases and the phase  separated states. It is shown that 
the $d$-wave pairing correlations are enhanced within the axial striped 
and phase separated states, but not in the regular phases. Moreover, it was 
found that the antiferromagnetic spin arrangement within the chains further 
enhances the $d$-wave paring correlations, while the ferromagnetic one has 
a fully opposite effect. This fact indicates that the type of spin ordering
plays an important role in the mechanism of stabilization of superconductivity
in strongly correlated systems and thus we have decided, within the current
paper, to examine this phenomenon in more detail. 

%It should be noted that unlike the  above mentioned 
%works [11,22] that also showed a significant enhancement of the d-wave 
%pairing correlations in striped phases introduced externally, in our 
%model the charge/spin stripes are present intrinsically and correspond 
%to  degrees of freedom in the actual materials. 

\section{Model}
The Hamiltonian of the model considered in this paper has the form  
\begin{eqnarray}
H&=&-t\sum_{\langle i,j\rangle\sigma}d^+_{i\sigma}d_{j\sigma} 
+ U_{fd}\sum_{i\sigma\sigma'}f^+_{i\sigma}f_{i\sigma}d^+_{i\sigma'}d_{i\sigma'}
+ J_z\sum_{i\sigma}(f^+_{i-\sigma}f_{i-\sigma} - 
f^+_{i\sigma}f_{i\sigma})d^+_{i\sigma}d_{i\sigma} 
\nonumber\\
&+& U_{dd}\sum_i d^+_{i\uparrow}d_{i\uparrow}d^+_{i\downarrow}d_{i\downarrow} \ ,
\end{eqnarray}
where $f^+_{i\sigma}, f_{i\sigma}$ are the creation and annihilation 
operators for an electron of spin $\sigma=\uparrow, \downarrow$ in the 
localized state at lattice site $i$ and $d^+_{i\sigma}, d_{i\sigma}$ are 
the creation and annihilation operators of the itinerant electrons in 
the $d$-band Wannier state at site~$i$.

The first term of (1) is the kinetic energy corresponding to quantum-mechanical 
hopping of the itinerant $d$ electrons between sites $i$ and $j$. These intersite
hopping transitions are described by the matrix elements $t_{ij}$, which are 
$-t$ if $i$ and $j$ are the nearest neighbours and zero otherwise.
The second term represents the 
on-site Coulomb interaction between the $d$-band electrons with density 
$n_d=N_d/L=\frac{1}{L}\sum_{i\sigma}d^+_{i\sigma}d_{i\sigma}$ and the localized
$f$ electrons with density  
$n_f=N_f/L=\frac{1}{L}\sum_{i\sigma}f^+_{i\sigma}f_{i\sigma}$, where $L$ is the 
number of lattice sites. The third term is the above mentioned anisotropic, 
spin-dependent local interaction of the Ising type between the localized 
and itinerant electrons that reflects the Hund's rule force. And finally,
the last term is the ordinary Hubbard interaction term for the itinerant 
electrons from the $d$ band. Moreover, it is assumed that the on-site Coulomb 
interaction between $f$ electrons is infinite and  so the double occupancy 
of $f$ orbitals is forbidden.

This model has several different physical interpretations
that depend on its application. As was already mentioned above, it can be
considered as the spin-one-half Falicov-Kimball model extended by the Hund 
and Hubbard interaction term. On the other hand, it can be also considered
as the Hubbard model in the external potential generated by the 
spin-independent Falicov-Kimball term and the anisotropic
spin-dependent Hund term. Very popular interpretation of the model
Hamiltonian (1) is its $(U_{dd}=0)$ version that has been introduced by 
Lemanski~\cite{Lemanski2} who considered it as the minimal model of charge and 
magnetic ordering in coupled electron and spin systems. Its attraction 
consists in this that without the Hubbard interaction term $(U_{dd}=0)$ 
the Hamiltonian (1) can be reduced to the single particle Hamiltonian
\begin{equation}
H=\sum_{ij\sigma}h^{(\nu)}_{ij}d^+_{i\sigma}d_{j\sigma},
\end{equation}
where $h^{(\nu)}_{ij}=t_{ij}+(U_{fd}w_i+J_z\nu{s_i})\delta_{ij}$,
$w_i=w_{i\uparrow}+w_{i\downarrow}=0,1$,
$s_i=w_{i\uparrow}-w_{i\downarrow}=-1,1$ 
and $\nu=\pm 1$. 
Thus for a given $f$-electron  $w=\{w_1,w_2,\dots,w_L\}$
and spin configuration $s=\{s_1,s_2,\dots,s_L\}$
the investigation of the model (2) is reduced to the investigation of the 
spectrum of $h^{(\nu)}$ for different $f$ electron/spin distributions. 
This can be performed exactly, over the full set of $f$-electron/spin 
distributions or approximatively. Numerical solutions obtained within
so called restricted set phase diagram method~\cite{Lemanski2,wrzodak} as wel as 
our well controlled gradient method~\cite{epjb,pssb} showed that this model is able 
to describe various types of charge and spin orderings observed 
experimentally in strongly correlated systems, including the diagonal 
and axial charge stripes with the antiferromagnetic or ferromagnetic arrangement 
of spins within the lines. Moreover, using the exact diagonalization
calculations~\cite{pssb} on small clusters ($L=16$) and the Projector 
Quantum-Monte-Carlo Method~\cite{epl} on larger clusters ($L \leq 64$), 
we have found that in the strong coupling $U_{fd}$ limit 
($U_{fd} \ge 4$) the ground states of the model (1) found for 
$U_{dd}=0$ persist as ground states also for nonzero $U_{dd}$, up to 
relatively large values ($U^c_{dd}\sim3$). This fact allows us to avoid 
the exhaustive numerical calculations on the full model Hamiltonian (1) 
and represent its ground states directly by a set of ground states of 
numerically much simpler single particle Hamiltonian (2), at least 
in the strong coupling $U_{fd}$ limit and $U_{dd} < U^c_{dd}$. 
For these ground states we then calculate the superconducting 
correlation functions of the full model Hamiltonian with $U_{dd} >0$
by the Projector Quantum-Monte-Carlo Method~\cite{PQMC}.

In particular we calculate the superconducting correlation function 
with $d_{x^2-y^2}$ wave symmetry defined as~\cite{Fettes}

\begin{equation}
C_d(r)=\frac{1}{L}\sum_{i,\delta,\delta'}g_{\delta}g_{\delta'}\langle
d^+_{i\uparrow}d^+_{i+\delta\downarrow}d_{i+\delta'+r\downarrow}d_{i+r\uparrow}
\rangle,
\end{equation}
where the factors $g_{\delta},g_{\delta'}$ are 1 in x-direction and
-1 in y-direction and the sums with respect to $\delta,\delta'$ are
independent sums over the nearest neighbors of site $i$.

However, on small clusters the above defined correlation function 
is not  a good measure for superconducting correlations,
since contains also contributions from the one particle 
correlation functions 
\begin{equation}
C^{\sigma}_0(r)=\frac{1}{L}\sum_{i}\langle d^+_{i\sigma}d_{i+r\sigma}
\rangle,
\end{equation}
that yield nonzero contributions
to $C_d(r)$ even in the noninteracting case. 

For this reason we use as the true measure for superconductivity
the vertex correlation function
\begin{equation}
C^{v}_d(r)=C_d(r)-\sum_{\delta,\delta'}g_{\delta}g_{\delta'}
C^{\uparrow}_0(r)C^{\downarrow}_0(r+\delta-\delta')
\rangle,
\end{equation}

and its average

\begin{equation}
C^{v}_d=\frac{1}{L}\sum_{i}C^{v}_d(i).
\end{equation}

\section{Results and discussion}
As mentioned above, the main goal of the present paper is to 
investigate the influence of the spin ordering on the superconducting
correlations in the ground state of the model Hamiltonian (1).
To fulfill this goal we have performed exhaustive numerical studies 
of the model for two selected values of $f$-electron fillings $N_f$
($N_f=L/2$ and $N_f=L$) and the complete set of even  $d$-electron 
filings $N_d$ on the cluster of $L=8 \times 8$ sites. The reasons 
for such a selection of $N_f$ values are following. The previous 
numerical results~\cite{pssb} obtained for the case $N_f=L/2$ showed that 
the ground states of the model (2) in this case are mainly the segregated 
or axial striped charge phases.  However, according to our very recent
results~\cite{epl} these configuration types enhance the $d$-wave pairing
correlations in the $d_{x^2-y^2}$ channel of the full model Hamiltonian
(1) and thus they are ideal candidates for the examination of effects 
of spin ordering on this charge induced superconducting state. 
In addition, it was found~\cite{pssb} that for both charge phases,
the segregated one as well as the  axial striped one, there are many different 
spin arrangements that minimize the ground state energy of the model 
at different $d$-electron fillings $N_d$. Thus it is possible to study 
simultaneously (by changing only one parameter $N_d$) the influence of 
the spin ordering and the $d$-electron doping on superconducting 
correlations. On the other hand the case of $N_f=L$ is of special 
importance for this reason that in this limit our model reduces on 
a simple spin-fermion model with an additional $U_{dd}$ interaction. 

Let us first discuss results obtained for $N_f=L/2$. In Fig.~1 
and Fig.~2 we present typical examples of ground states, that
minimize the ground state energy of the model Hamiltonian (2) for
$U_{fd}=4$,  $J_z=0.5$ and $U_{dd}=0$. 
\begin{figure}[h!]
\begin{center}
\includegraphics[width=16cm]{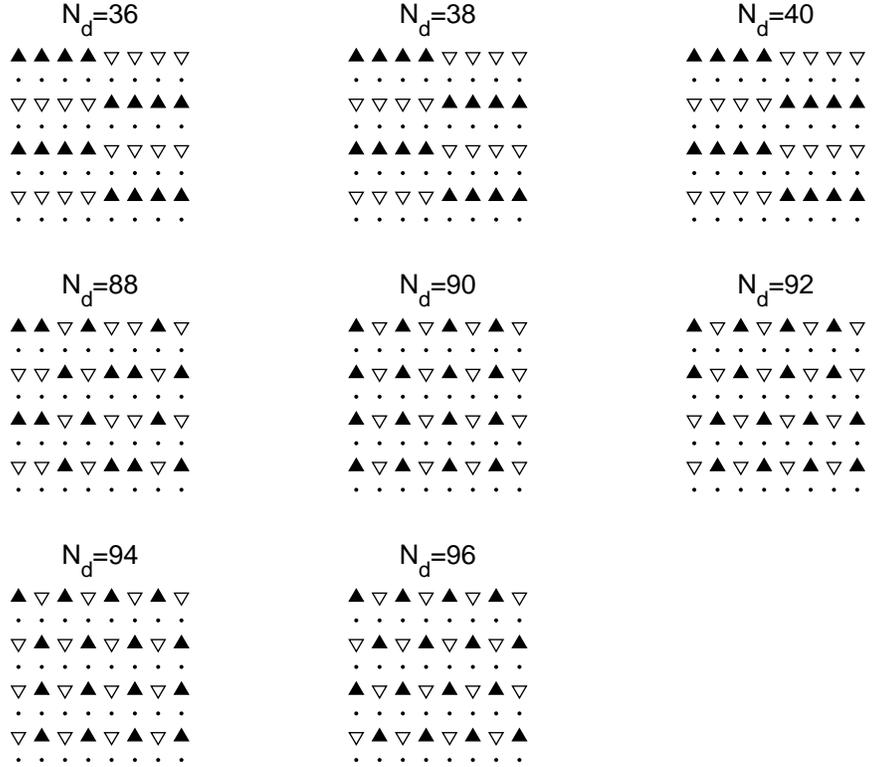}
\end{center}
\caption{\small The axial striped ground states of the model (1) obtained for $U_{fd}=4,
J_z=0.5, U_{dd}=0$ and  $N_f=L/2$ on the $L=8 \times 8$ site cluster.   
Here the spin up (down) of the $f$ electron is represented by a filled 
regular triangle (open inverted triangle).
}
\label{fig1}
\end{figure}
\begin{figure}[h!]
\begin{center}
\includegraphics[width=16cm]{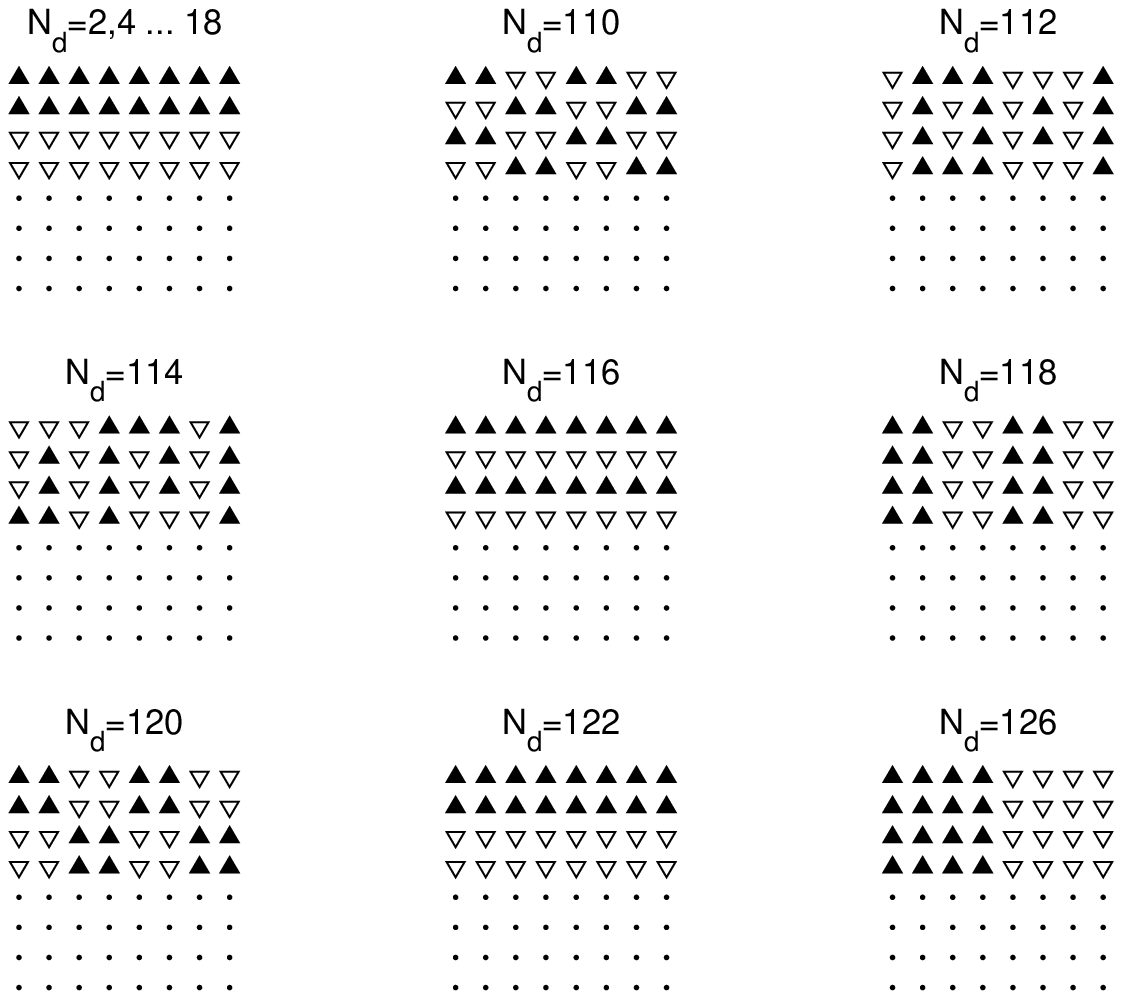}
\end{center}
\caption{\small 
The phase segregated ground states of the model (1) obtained for $U_{fd}=4,
J_z=0.5, U_{dd}=0$ and  $N_f=L/2$ on the $L=8 \times 8$ site cluster.
}
\label{fig1}
\end{figure}
One can see that for both the axial striped
as well as phase segregated phase there are several different 
spin arrangements that allow us to test the impact of spin ordering 
on the superconducting correlations. Before this let us discuss
in more detail these configurations types. In all examined cases the ground 
states of the model are non-polarized ($S_z=0$) 
for both the axial striped and segregated phase. For the axial striped phase
the one dimensional chains are formed by (i) the four-spin ferromagnetic 
clusters of opposite orientation, (ii) the mixture of two-spin ferromagnetic
clusters and $\uparrow \downarrow$ or  $\downarrow \uparrow$ pattern and 
(iii) the classical Neel state pattern $\uparrow \downarrow \dots  \uparrow \downarrow$.    
A similar situation we can observe also in the phase segregated phase.
Here we can find (i) the antiparallel ferromagnetic chains (bands), (ii)
the antiparallel small or large ferromagnetic domains and (iii) some 
intermediate phases.

The average vertex correlation functions $C^{v}_d$ corresponding to these 
ground states are displayed in Fig.~3 (for the axial striped phase) and 
Fig.~4 (for the segregated phase).
\begin{figure}[h!]
\begin{center}
\includegraphics[width=14cm]{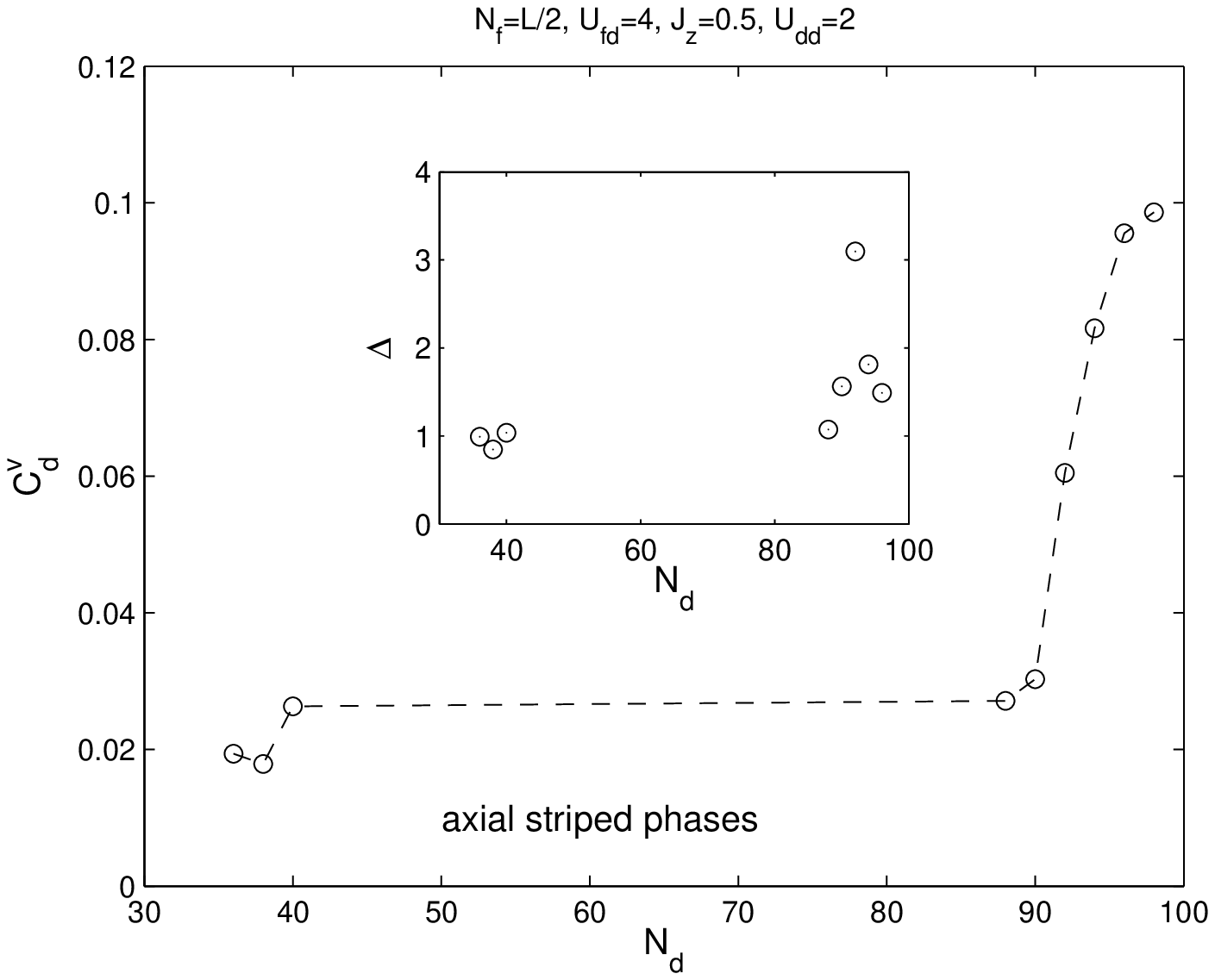}
\end{center}
\caption{\small Average vertex correlation function $C^v_d$ with 
$d_{x^2-y^2}$-symmetry calculated for corresponding ground states from 
Fig.~1. The inset shows the enhancement $\Delta$ corresponding to the ratio 
of the average vertex correlation functions with and without the Ising 
coupling $J_z$, $\Delta=C^v_d(J_z=0.5)/C^v_d(J_z=0)$.
}
\label{fig1}
\end{figure}
\begin{figure}[h!]
\begin{center}
\includegraphics[width=14cm]{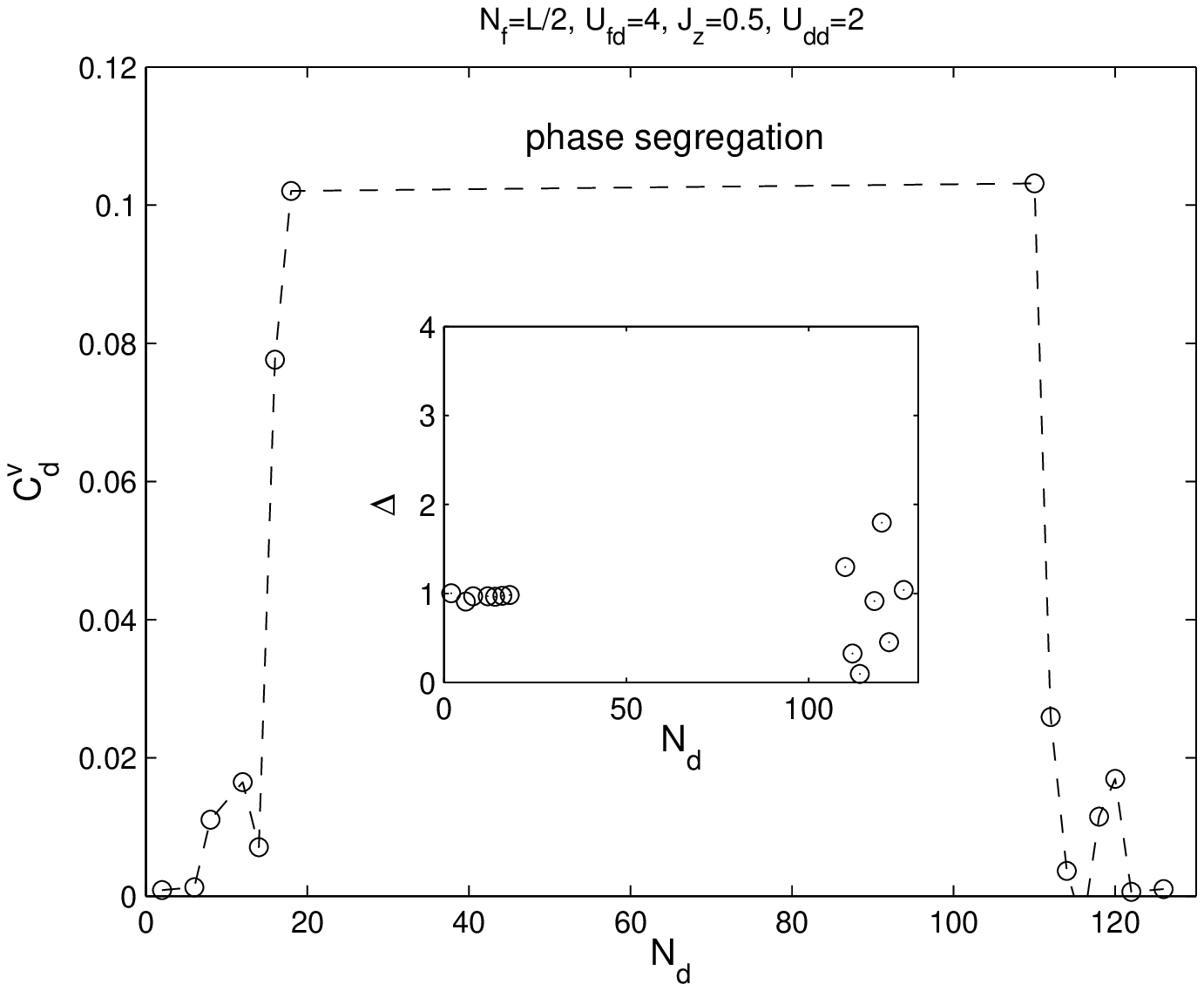}
\end{center}
\caption{\small 
Average vertex correlation function $C^v_d$ and the enhancement $\Delta$ 
calculated for corresponding ground states from Fig.~2. 
}
\label{fig1}
\end{figure}
One can see that in the case of axial
striped phases the superconducting correlations are enhanced the most
significantly for the chessboard distribution of spins. All deviation from this 
state in the meaning of forming the ferromagnetic clusters or an improper
extension of the chessboard structure in the y-direction suppress the superconducting 
correlations in the $d$-wave channel. Moreover, there is observed 
an obvious relation between the size of ferromagnetic clusters (domains)
and the superconducting correlations, the largest ferromagnetic clusters,
the smallest vertex correlations. However, the vertex correlations in 
Fig.~3 are displayed for different $d$-electron fillings and thus it is 
questionable if the enhancement/suppression of vertex correlations is a net
effect of different spin orderings, or it is, at least partially, produced by
the $d$-electron doping. To separate contributions to $C^v_d$ from $N_{d}$ and 
$J_z$ we have plotted in the inset to Fig.~3 the ratio of the average vertex
correlation functions with and without the Ising coupling $J_z$, 
$\Delta=C^v_d(J_z=0.5)/C^v_d(J_z=0)$.
These results show that for $N_d=36,38,40$, where the ground states
are identical, the ratio $\Delta$ depends only very weakly on $N_d$
what documents that the effects of electron doping on superconducting
correlations are not very important. On the other hand the results 
from the opposite limit $N_d>L/2$ show the strong enhancement of $\Delta$ 
in the region where the ground states are different non-polarized 
spin orderings without or with small ferromagnetic clusters of length two, 
what clearly demonstrates the impact of such  a spin ordering on the 
superconducting correlations.  

A slightly different behaviour of the model is observed in the segregated phase
(see Fig.~4). Here the superconducting correlations are strongly enhanced 
going with $N_d$ from 2 to 18. Since the spin ordering in all these cases is
identical, the enhancement of superconducting correlations in this region 
is obviously a net effect of electron doping. In the opposite limit $N_d>L/2$   
the superconducting correlations are enhanced for the smallest 
ferromagnetic clusters and they are strongly suppressed with the 
increasing size of ferromagnetic clusters (domains). In this 
region the increase (decrease) in $C^v_d$ does not fully coincides
with increase (decrease) in $\Delta$ what indicates that the 
enhancement of the superconducting correlations for $N_d > L/2$
is the  combined effect of spin ordering and the $d$-electron doping.  

Let us now turn our attention to the case $N_f=L$. The ground states of
the single particle Hamiltonian (2), that are used as the approximative
ground states of the full model Hamiltonian (1) are displayed in Fig.~5.
\begin{figure}[h!]
\begin{center}
\includegraphics[width=16cm]{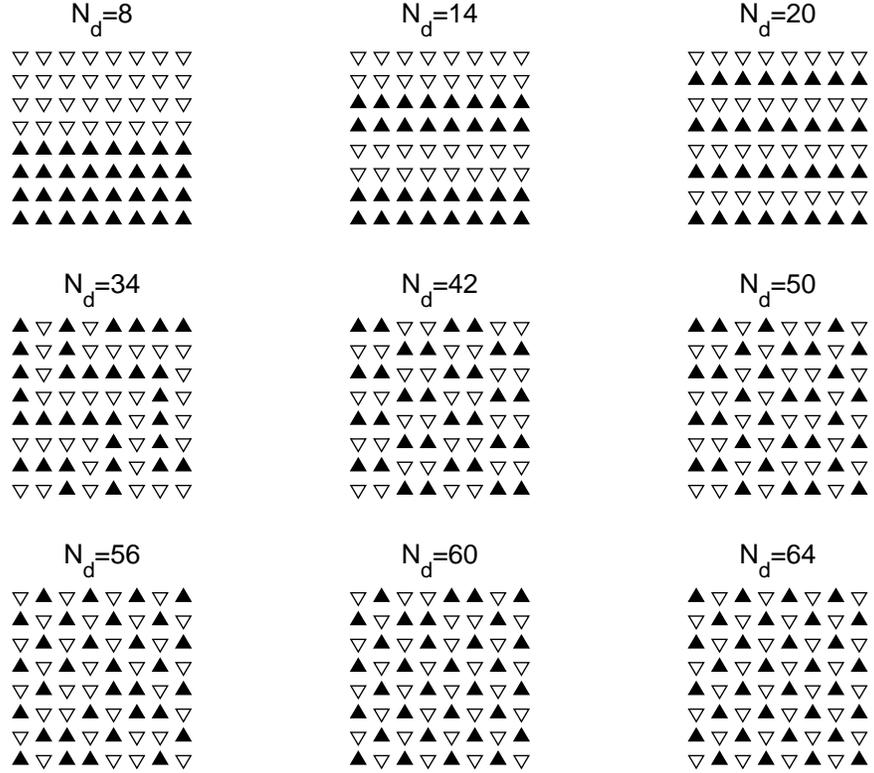}
\end{center}
\caption{\small 
Ground states of the model (1) obtained for $U_{fd}=4,
J_z=0.5, U_{dd}=0$ and  $N_f=L$ on the $L=8 \times 8$ site cluster.
}
\label{fig1}
\end{figure}
Obviously there are some general trends in the spin ordering going with $N_d$
from 0 to L. For $N_d$ small (e.g., $N_d=8$) the ground state is formed 
by two large antiparallel ferromagnetic domains, that transforms with 
increasing d-electron filling $N_d$ on antiparallel ferromagnetic bands 
(e.g., $N_d=14$) and finally on antiparallel ferromagnetic chains 
(e.g., $N_d=20$). Then follows the region of perturbed  antiparallel 
ferromagnetic chains (e.g., $N_d=34$) and the region of regularly 
distributed pairs of  up and down spins (e.g., $N_d=42$). The next phases 
can be considered as a mixture of this regular phase and the chessboard 
phase (e.g., $N_d=50$). Then follows the region of incompletely developed 
chessboard phase  (e.g., $N_d=56$), which ends with the prefect developed
chessboard structure at $N_d=L/2$.     

The average vertex correlation functions $C^{v}_d$ corresponding to these 
ground states are displayed in Fig.~6. 
\begin{figure}[h!]
\begin{center}
\includegraphics[width=14cm]{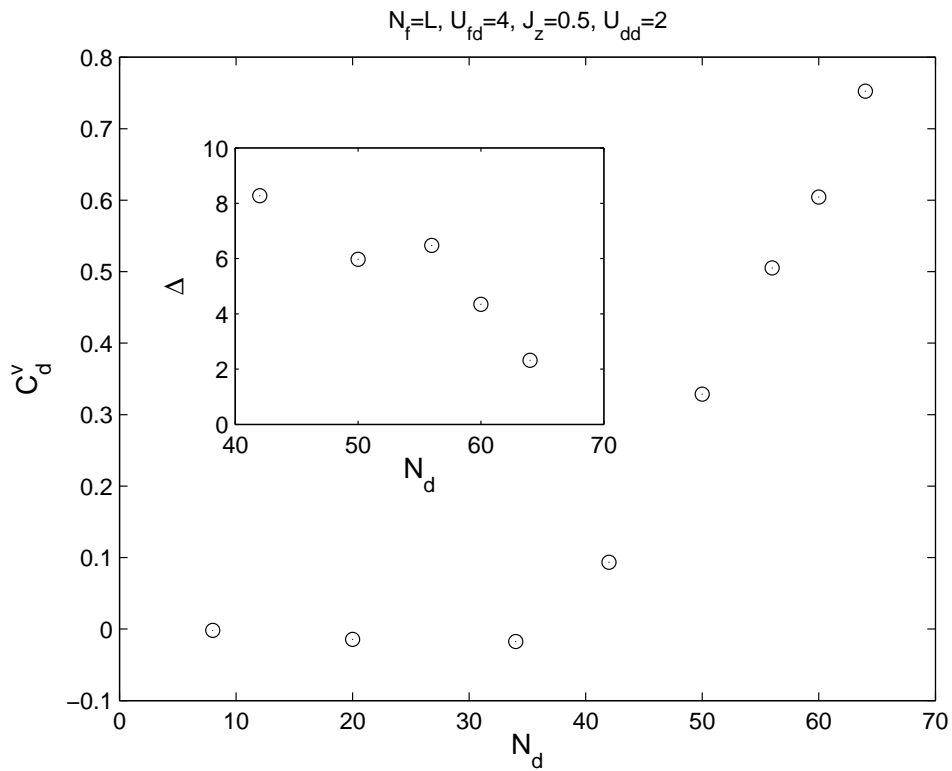}
\end{center}
\caption{\small 
Average vertex correlation function $C^v_d$ and the enhancement $\Delta$ 
calculated for corresponding ground states from Fig.~5. 
}
\label{fig1}
\end{figure}
It is seen that the superconducting 
correlations are negligible in phases that are composed of large antiparallel 
ferromagnetic domains, bands and chains what is fully consistent with 
our above discussed results and conclusions. The superconducting
correlations start to increase from the region of stability of 
regular phases, what is also in accordance with our above mentioned 
conclusions, since these phases are formed by small antiparallel 
ferromagnetic clusters of length two. In the region of formation
of the chessboard structure, the vertex correlation function 
is dramatically enhanced and reaches its maximum for the perfect 
ordered chessboard phase (for $N_d > L/2$ the vertex correlation function
$C^v_d$ exhibits the mirror symmetry). The same behaviour exhibits also the 
ratio $\Delta$ that is enhanced by factor 6-8 in comparison to the 
$J_z=0$ case, what clearly documents the strong effects of spin ordering on
superconducting correlations in coupled electron and spin systems.

In summary, we have used the generalized spin-one-half Falicov-Kimball model 
with Hund and Hubbard coupling to study effects of spin ordering on 
superconducting correlations in the axial striped and phase
segregated state. It was found that the ferromagnetic spin clusters
(lines, bands, domains) suppress the superconducting correlations
in the d-wave chanel, while the antiferromagnetic ones have 
the fully opposite effect. The enhancement of the superconducting
correlations due to the antiferromagnetic  spin ordering is by factor 3
in the axial striped phase and even by the factor 8 in the phase segregated 
phase.

\vspace{0.4cm}
This work was supported by Slovak Research and Development Agency (APVV)
under Grant APVV-0097-12 and ERDF EU Grants under the contract No.
ITMS 26220120005 and ITMS26210120002.

\end{document}